\documentclass[11pt,a4paper]{article}

\usepackage[T1]{fontenc}
\usepackage[margin=1in]{geometry}
\usepackage{cite}
\usepackage{amsmath,amssymb,amsfonts,amsthm}
\usepackage{graphicx}
\usepackage{textcomp}
\usepackage{booktabs}
\usepackage{bm}
\usepackage{xurl}
\usepackage[hidelinks]{hyperref}

\hypersetup{
  pdftitle={A Matched Evaluation of QUBO Action Models and Defensive Remediation in Cyber Attack Graphs},
  pdfauthor={Kasymkhan Khubiev, Yaroslav Kholodov, and Alexey S. Petrenko}
}

\newcommand{\Aset}{\mathcal{A}}
\newtheorem{definition}{Definition}
\newtheorem{proposition}{Proposition}

\title{A Matched Evaluation of QUBO Action Models and Defensive
Remediation in Cyber Attack Graphs}
\author{
  Kasymkhan Khubiev\thanks{Corresponding author: \href{mailto:kasymkhankhubievnis@gmail.com}{kasymkhankhubievnis@gmail.com}.}$^{1}$
  \and Yaroslav Kholodov$^{1}$
  \and Alexey S. Petrenko$^{2}$
}
\date{}

\begin{document}
\maketitle

\begin{center}
\small
$^{1}$Sirius University of Science and Technology,\\
1, Olimpiyskiy Ave., Sirius, Sirius Federal Territory,\\
Krasnodar region, 354340, Russia\\
\href{mailto:kasymkhankhubievnis@gmail.com}{kasymkhankhubievnis@gmail.com};
\href{mailto:kholodov.ya@talantiuspeh.ru}{kholodov.ya@talantiuspeh.ru}
\par\medskip
$^{2}$St. Petersburg Federal Research Center of the Russian Academy of Sciences,\\
39, 14th Line V.O., St. Petersburg, Russia, 199178\\
\href{mailto:a.petrenko1999@rambler.ru}{a.petrenko1999@rambler.ru}
\end{center}

\begin{abstract}
This study examines how vulnerability-specific actions affect quadratic
unconstrained binary optimization (QUBO) models for minimum-cost cyber attack
paths and defensive remediation. Edge-flow and variable-length hop
formulations are evaluated using simulated annealing, steepest descent, and
tabu search against exact classical references. Experiments on six independent
synthetic topologies separate the effects of transition eligibility,
vulnerability labels, exploit costs, and multiple exploit alternatives. A
single labeled action per eligible transition preserves feasible paths,
optimal costs, and QUBO size. Under the tested settings, multiple exploit
alternatives were associated with approximately unchanged valid-path recovery,
lower optimal-path recovery, and substantially larger representations,
particularly for the hop formulation. These recovery differences remain
descriptive because they are
not statistically significant after correction for multiple comparisons. An
evaluation on a published 3CX attack graph with synthetic costs demonstrates
applicability beyond the generated topologies. The defensive extension selects
patches within a budget to raise attack-path costs to specified thresholds or
disconnect protected targets. Against exhaustive and mixed-integer references,
the tested QUBO heuristics recovered an optimal patch set in 12 of 16 calls
across eight scenarios on one topology.
The results characterize formulation correctness, solution recovery, and
resource requirements using classical local optimization, without establishing
quantum computational advantage.
\end{abstract}


\section{Introduction}
\label{sec:introduction}

Modern society depends on interconnected enterprise services, public
infrastructure, Internet of Things devices, financial platforms, and
blockchain-based systems. A vulnerability in one component can therefore
become an entry point to a wider compromise: loss of confidentiality,
disruption of essential services, manipulation of transactions, and lateral
movement into otherwise protected assets.  Such attacks often unfold as
ordered, multi-step processes rather than isolated events, while
resource-constrained networks also face the practical problem of tracing malicious
traffic back through several relays~\cite{konsta2024,antwi2025}.  Preventive
security consequently requires more than detecting a single exposed host; it
requires identifying which combinations of reachability and vulnerability can
carry an attacker from an entry point to a high-value target.

Several complementary approaches address this need.  Vulnerability assessment
and patch management identify and remove exposed weaknesses; intrusion
detection and traceback observe malicious behavior; attack trees and attack
graphs represent the prerequisites and alternatives of a multi-stage
compromise; and path-extraction methods prioritize the chains that deserve
remediation or monitoring~\cite{konsta2024,levner2024}.  This paper begins
after network and vulnerability discovery: it assumes that a directed graph
and an applicable action catalog have been constructed, then studies how a
minimum-cost entry-to-target chain can be recovered reproducibly.

A useful controlled abstraction is the positive-weight shortest-path problem.
Systems are vertices, allowed movements are directed transitions, and each
transition receives an additive cost representing the analyst's chosen notion
of attacker effort.  For non-negative costs, Dijkstra's
algorithm~\cite{dijkstra1959} supplies an exact solution efficiently.  Ordinary shortest path is
therefore a methodological baseline for formulation and stochastic recovery;
it does not by itself establish a compelling application for demonstrating
quantum computational advantage.

To examine how vulnerability information changes this abstraction, we first
formulate the shortest-path problem using only network connectivity and edge
costs.  This network-only formulation, denoted \textbf{C1}, serves as the
reference for the vulnerability-action model introduced below.  Its decision
variables select network edges $(u,v)$, and its objective is the sum of their base network
weights.  C1 provides a baseline for quadratic unconstrained binary
optimization (QUBO) construction, optimization, path decoding,
and comparison against an exact reference.
At this level of abstraction, an edge specifies an allowed network transition,
while the vulnerability enabling the next compromise remains implicit.

To make this vulnerability information explicit, we extend C1 to a
vulnerability-action model, denoted \textbf{C2}.  Each eligible network edge is
represented by one or more actions $a=(u,v,k)$, where $k$ identifies the
vulnerability used to compromise $v$ from $u$.  Each action has a scalarized positive cost incorporating the
network transition and exploit-related components.  C2 therefore asks for an
ordered route-and-vulnerability chain rather than only a route.  In the
present unconstrained baseline, it remains exactly solvable by Dijkstra on the
resulting action multigraph.  We use that path problem as a controlled
formulation baseline and then study threshold remediation, where one patch can
disable actions on several paths and protected pairs. Dijkstra then identifies
residual attack paths that violate the defender's thresholds, allowing
additional constraints to be generated. Logical
AND/OR prerequisites remain outside the present model.

QUBO expresses both objectives
as quadratic polynomials over binary
variables~\cite{glover2019,lucas2014}.  The same model can be submitted to
classical heuristics, quantum-inspired optimizers, hybrid methods, or quantum
annealers.  An optimizer nevertheless returns a binary sample rather than a
certified attack path.  A security result must decode to one continuous
source-to-target chain, select one applicable action for each compromise, and
match the independently computed minimum cost whenever exact recovery is
claimed.

Quantum computing and physics-inspired Ising machines provide emerging
paradigms for optimizing such energy models.  Quantum annealing seeks
low-energy configurations through an adiabatic evolution of a physical Ising
system.  Simulated bifurcation instead numerically evolves classical nonlinear
Hamiltonian systems whose bifurcating states encode Ising spins; its
simultaneous updates make it amenable to massively parallel
computation~\cite{goto2019,goto2021}.  Comparative benchmarks show that ballistic and
discrete simulated-bifurcation variants can be competitive with quantum
annealers on particular Max-Cut graph families, while also demonstrating that
performance depends on topology, target quality, and the chosen success
metric~\cite{zeng2024}.  These methods motivate a common QUBO/Ising
representation, but none removes the need for application-specific decoding
and an exact classical reference.

Quantum computing also motivates security priorities through its potential
cryptanalytic capabilities. The Quantum Risk Index (QRI) proposed by
Ishchukova et al.~\cite{ishchukova2026} combines Common Vulnerability Scoring
System (CVSS) severity with susceptibility to quantum algorithms and
harvest-now-decrypt-later exposure. Such scores could inform protection
priorities in a defensive optimization model, but they measure security risk
rather than optimizer performance. A higher risk score also does not imply
greater attacker effort and should not automatically become a higher cost in
a minimum-cost attack-path objective. The present experiments use synthetic
action and patch costs without QRI inputs, and evaluate classical QUBO
heuristics rather than quantum cryptanalysis or physical quantum hardware.

The empirical question is how eligibility, labels, exploit costs, and action
multiplicity separately change QUBO recovery and resources.  The experiment
pairs all five stages on the same master topologies and endpoints, compares
edge-flow and variable-length hop formulations under three local classical
QUBO heuristics, and separates successful calls from their nested raw reads.
Dijkstra remains the exact path benchmark.  Exhaustive remediation-set
enumeration and a constraint-generating mixed-integer model independently
certify the coupled defender extension.

The paper makes five contributions:

\begin{itemize}
    \item a matched sequence that isolates eligibility, labels,
    exploit costs, and action multiplicity;
    \item sufficient-penalty QUBOs for action edge-flow and ordered action-hop
    models, including safe presolve and primary-cost-preserving tie breaking;
    \item an exploratory experiment over six graph sizes, two nested densities,
    four source--target pairs per topology--density instance selected using
    construction and difficulty-based policies, three backends, and two formulations with
    topology-clustered inference;
    \item descriptive analysis of recovery patterns, penalty sensitivity, and performance
    on a published attack graph;
    \item a budgeted threshold-remediation QUBO with exhaustive and iterative
    mixed-integer reference solutions.
\end{itemize}

Section~\ref{sec:previous-work} connects attack-path analysis to classical shortest path, QUBO, and
Ising-machine research.  Sections~\ref{sec:research-questions}--\ref{sec:methodology} then state the research questions and
define the paired data, C1/C2 objectives, formulations, decoders, and
inferential design. Section~\ref{sec:results} reports matched-stage recovery, resources,
external evaluation, sensitivity, and remediation.  Sections~\ref{sec:discussion}--\ref{sec:reproducibility} interpret the
results, discuss limitations, and describe the reproducibility materials.

\section{Previous Work and Literature Overview}
\label{sec:previous-work}

\subsection{Cyber Attacks, Attack Graphs, and Ordered Paths}

Attack trees and attack graphs are established representations of multi-step
security scenarios.  Konsta et al.~\cite{konsta2024} classify automatic generation approaches
as model-driven, analysis-driven, or vulnerability-driven and emphasize that
the provenance and semantics of the resulting graph matter.
The survey has a published corrigendum~\cite{konsta2024corr}.
Their survey also defines an important boundary for this work: we evaluate
path recovery on generated graphs but do not claim to automate attack-graph
construction.

Levner and Tsadikovich~\cite{levner2024} describe an attack path as an ordered sequence of
attacker actions and formulate extraction as graph optimization with actions
on nodes and edges.  Their Dijkstra-type method handles a
cyclic AND/OR attack-graph class.  Its use of graph optimization connects attack-path extraction to the present
study, although C2 represents positive-weight action chains without the logical
prerequisites of an AND/OR attack graph.

\subsection{Shortest Path and QUBO Foundations}

For a positive-weight, single-source, single-target problem, Dijkstra is the
appropriate exact reference.  Bellman--Ford~\cite{bellman1958,ford1956}
supports negative weights provided that no reachable negative cycle affects
the source--target path cost. A*~\cite{hart1968,hart1972corr} incorporates an admissible
heuristic, and Floyd--Warshall~\cite{floyd1962} computes all pairs; these settings are not
the present experimental target.  Because specialized polynomial algorithms
already solve C1 and unconstrained C2, the role of QUBO here is to expose
representation and recovery behavior under controlled conditions.

In a QUBO, constraints are folded into an energy function using penalties.
Lucas's catalog~\cite{lucas2014} illustrates the central requirements: decision-variable
meaning, sufficient penalty dominance, and resource growth must be justified
alongside the cost objective.  A valid Ising or QUBO mapping
does not imply that an annealer solves the encoded problem efficiently, and
low energy does not by itself establish path validity.

Krauss and McCollum~\cite{krauss2020} provide the principal C1 formulation precedent.  Their
directed edge-flow model uses one variable per edge, whereas their
vertex-position model makes order explicit at a larger variable cost and
requires a path-length convention.  These alternatives
motivate the present edge-flow/hop comparison. The present evaluation combines
strict decoding, explicit terminal conditions, sufficient-penalty arguments,
exact classical references, and paired statistical analysis.

Rovara, Quetschlich, and Wille~\cite{rovara2026} automate the construction of
pathfinding QUBOs from composable domain-level constraints and permit one-hot,
domain-wall, and binary encodings to be exchanged without manually rebuilding
the formulation.  Their case studies demonstrate that the most
resource-efficient encoding is problem-dependent and that encoding choice
changes the valid-state landscape.  The present work has a narrower but
complementary goal: it derives and evaluates two cybersecurity-specific
action-path formulations and measures their valid and optimal stochastic
recovery over repeated paired graph instances.

\subsection{Quantum Annealing and Simulated Bifurcation}

The QUBO--Ising equivalence opens several solver routes.  Quantum annealers
physically implement a time-dependent Ising Hamiltonian and seek a low-energy
final state through quantum adiabatic evolution.  Their practical performance
is conditioned by connectivity, embedding, noise, coefficient scaling, and
sampling effort; an Ising mapping alone is not evidence of an advantage.

Goto, Tatsumura, and Dixon~\cite{goto2019} introduced simulated bifurcation as a classical
heuristic inspired by quantum adiabatic optimization with nonlinear
oscillators.  Its two bifurcation branches encode binary spin
states, and simultaneous variable updates support parallel implementations
on field-programmable gate arrays (FPGAs) and graphics processing units (GPUs).
Later ballistic and discrete variants were designed to
reduce analog errors and improve solution accuracy~\cite{goto2021}.  Zeng
et al.~\cite{zeng2024} benchmark these and other quantum-annealing-inspired algorithms
against D-Wave quantum annealing on Max-Cut instances and show that rankings
depend on graph family, size, and whether success probability, time to
solution, or time to target is measured.  Simulated
bifurcation is therefore a relevant future backend and a useful baseline for
quantum annealing, but it is not evaluated in the present study.

\subsection{Empirical and Application-Adjacent Evidence}

Oliva-Gonz\'alez and Jim\'enez-Hern\'andez~\cite{oliva2026} evaluate simulated annealing for
the network shortest-path problem corresponding to our C1 baseline
on controlled synthetic graphs and a 443-node urban network, reporting success
probability and confidence-based time to solution.  Their
reported recovery trends and reliability--runtime trade-off
motivate the present separation of a useful call from an individual read.
Singh et al.~\cite{singh2025} apply one shortest-path Hamiltonian across the
quantum approximate optimization algorithm (QAOA), simulated
quantum annealing, and adiabatic simulation on small instances, illustrating
the sensitivity of recovery to backend and schedule choices while also showing
why simulator behavior must be separated from physical hardware
performance.

Cybersecurity path reconstruction has appeared in adjacent QUBO work.  Antwi
et al.~\cite{antwi2025} formulate Internet Protocol (IP) traceback in wireless
sensor networks through link
selection and path-continuity penalties.  The task differs
from C2: traceback reconstructs packet provenance, whereas C2 identifies the
vulnerability action enabling each compromise.  Nevertheless, it reinforces
the need to report invalid reconstructions and hardware-specific embedding
costs rather than energy alone.

Carney~\cite{carney2022} provides a direct predecessor for QUBO-based
defensive prioritization.  That study maps host--vulnerability incidence to an
undirected vulnerability-dual graph and solves a minimum vertex-cover QUBO on
D-Wave hardware to disrupt vulnerability co-occurrence chains. The study also
discusses weighted minimum vertex cover. Our remediation model addresses
directed threshold path hitting with positive attack costs and explicit patch
budgets, coupling vulnerabilities shared across actions and protected pairs.
Both approaches accommodate nonuniform costs. The present evaluation uses
strict semantic decoding and independent exhaustive and mixed-integer
references.

Ishchukova et al.~\cite{ishchukova2026} propose a QRI
that augments CVSS severity with keyword-derived factors representing
susceptibility to Shor's and Grover's algorithms and exposure to
harvest-now-decrypt-later threats.  Applied to 78,587 National Vulnerability
Database (NVD) records, the index was associated with membership in the
Cybersecurity and Infrastructure Security Agency (CISA) Known Exploited
Vulnerabilities catalog and provided a quantum-aware triage dimension
complementary to the Exploit Prediction Scoring System (EPSS).  Their study
focuses on individual records in the Common Vulnerabilities and Exposures
(CVE) catalog and does not model how
applicable vulnerabilities compose into ordered attack chains or coupled
remediation decisions.  The present work addresses this adjacent decision
layer by representing vulnerabilities as actions on directed paths and
optimizing patch sets across shared vulnerabilities and protected pairs.
QRI-related attributes could inform future protection priorities in the
remediation model.  Incorporating them would require an explicit mapping to
security consequences or protection requirements, separate from attacker
effort, exploitation likelihood, and patch cost.  This extension is not part
of the present experiments.

Richer routing problems show how added semantics affect representation.
Postman variants add coverage, ordering, capacity, and multiple-agent
constraints~\cite{pion2023}; ride-pooling QUBOs demonstrate that logically
equivalent encodings can have sharply different variable and coupling
costs~\cite{cattelan2024}.  These studies do not predict C1 or C2 recovery
rates, but they motivate the resource-aware comparison and the future move
toward defensive optimization with coupled constraints.

\subsection{Synthesis and Research Gap}

The literature establishes attack graphs as representations of multi-step
security behavior, shortest path as a tractable analytical baseline, and
edge-flow and position models as established QUBO encodings.  Together, these
foundations motivate a paired comparison that holds network
topology, endpoint, formulation, backend, and sampling effort fixed while
lifting network edges into vulnerability-labeled action choices.

This study realizes that comparison through a controlled and reproducible
experimental design.  Pairing the five semantic conditions on identical topologies and endpoints
isolates eligibility, label representation, exploit costs, and multiplicity
as adjacent changes.  The evaluation distinguishes structural validity from
exact optimality, separates read-level outcomes from call-level recovery,
clusters inference at the master-topology level, and quantifies the variable
and coupling overhead of action order.  Dijkstra provides the exact path
reference, while independent exhaustive and mixed-integer references certify
the threshold-remediation extension.

\section{Research Questions}
\label{sec:research-questions}

\begin{description}
  \item[RQ1.] \textbf{Representation parity.} Does a matched one-action representation
  preserve the eligible network's feasible paths, optimal costs, ground states
  under the edge-to-action variable correspondence, and QUBO resources?
  \item[RQ2.] \textbf{Semantic decomposition.} What are the separate paired effects
  of eligibility, labels, exploit costs, and action multiplicity on valid and optimal
  recovery and resources?
  \item[RQ3.] \textbf{Recovery patterns and resources.} How do optimal-path
  recovery and representation size vary with action multiplicity, formulation,
  and heuristic under the tested settings?
  \item[RQ4.] \textbf{External evaluation.} How accurately do the formulations recover
  valid and optimal paths on a published attack graph?
  \item[RQ5.] \textbf{Defensive remediation.} Can a QUBO recover an optimal patch set
  within a budget that blocks every sub-threshold chain across shared
  vulnerabilities and multiple protected pairs?
\end{description}

\section{Data and Experimental Design}
\label{sec:experimental-design}

The primary analysis uses a resource-pilot dataset with one independently
seeded master topology at each size $n\in\{6,8,10,14,20,24\}$. Each is
observed at two nested target densities, 0.20 and 0.35, while preserving
identities and attributes on shared edges; realized densities reflect integer
edge counts. Four source--target pairs are selected per topology--density
instance before QUBO sampling: one construction pair, two low-difficulty pairs,
and one high-difficulty pair. Difficulty combines unweighted path length,
feasible-path count, and the gap between the best and second-best path costs;
path enumeration is capped at 1,000 paths. This gives 48 nested
topology--density--pair observations. The master topology is the independent
inferential unit. The dataset is exploratory and is distinct from the larger
confirmatory protocol included in the repository.

Every observation passes through the nested semantic sequence in
Table~\ref{tab:matched-design}.  The multiple-action condition retains the
single exploit-cost action and adds two stable alternatives per eligible
transition.  For each condition, edge-flow and variable-length terminal-padded
hop formulations are evaluated using simulated annealing, steepest descent,
and tabu search.  A solver call is one execution of a heuristic on a QUBO;
a read is one returned binary sample.  Each primary-study call contains
10 reads; simulated
annealing uses 100 sweeps and a fixed geometric beta schedule over
$[0.1,4.0]$.  Maximum-absolute-bias normalization and sufficient penalties
are applied to both formulations. The shared solver seed is 101. Steepest
descent and tabu search use random initial states; tabu search uses a 10\,ms
timeout parameter. Equal read counts do not imply equal computational effort
across heuristics. All five semantic conditions use the same presolve,
action-count augmentation, constructive penalty policy, and decoder; the
Appendix specifies these shared rules for network and action variables.

\begin{table}[t]
\caption{Matched semantic sequence.  Adjacent differences isolate each
individual change.}
\label{tab:matched-design}
\centering
\small
\begin{tabular}{@{}lll@{}}
\toprule
Condition & Construction & Isolated interpretation \\
\midrule
Full network & all directed edges, network costs & network baseline \\
Eligible network & exploit-eligible edges, same costs & eligibility \\
Matched action & one labeled action per eligible edge & label representation \\
Exploit cost & one action with decomposed cost & cost semantics \\
Multiple action & three nested actions per eligible edge & action multiplicity \\
\bottomrule
\end{tabular}
\end{table}

The primary experiment uses a paired factorial design.  For each combination
of topology, density, and source--target pair, the five semantic conditions
are evaluated with both QUBO formulations and all three heuristics.  A common
random seed is used across these comparisons, and Dijkstra provides the exact
minimum path cost for each condition.  This pairing allows changes in recovery
and representation size to be compared on the same underlying network.

Three additional experiments assess penalty sensitivity, external
applicability, and remediation.  The sensitivity experiment compares four
penalty and sampling-schedule combinations on one topology.  The external
evaluation uses the 3CX case from version v1 of the published MulVAL
replication package by Soeiro~\cite{soeiro2025},
identified by digital object identifier (DOI) \texttt{10.5281/zenodo.15924457} and distributed
under the GNU General Public License (GPL), version 3 or later.
The evaluated graph is a conversion of the source case, retaining entity
identities, directed relation facts, and named vulnerabilities. Bridge
transitions inferred from the source narrative are marked explicitly in the
conversion artifacts; numeric costs and the two additional actions used for
multiplicity sensitivity are synthetic. Uniform,
relation-weighted, and high-detection synthetic cost scenarios are evaluated
for two targets, \texttt{h10\_victim} and \texttt{h11\_victim}, in 180 calls
with 20 reads each. The remediation study uses one six-node topology and
crosses observed versus deterministically reduced action availability, low
versus high vulnerability sharing, and one versus two protected pairs at
a budget of two patch-cost units, producing eight scenarios with exact
reference solutions and 16 QUBO calls. Reduced availability removes 20\% of
actions deterministically. Each remediation call returns 100 reads;
simulated annealing uses 1,000 sweeps, and the other tested heuristic is
steepest descent. Each protected pair's threshold is its initial shortest-path
cost plus one cost unit.
These experiments are analyzed separately from the primary path-recovery
experiment.

\section{Methodology}
\label{sec:methodology}

\subsection{Formulations and Path Decoding}

The edge-flow and hop formulations represent the same minimum-cost
source--target path problem through different binary encodings.  Edge-flow
selects network transitions or vulnerability actions subject to flow
conservation, whereas hop assigns them to successive positions in an ordered
chain.  Comparing these encodings reveals how the explicit representation of
path order affects solution recovery and model size.

To keep this comparison consistent, both formulations apply the same graph
reductions and derive their penalty strengths from a known feasible path.
The penalties make constraint violations more costly than that path, while
positive coefficient normalization preserves the energy ordering.  The hop
formulation uses a safe upper bound on path length and represents shorter
paths with inactive terminal positions, without adding to their cost.
Formal definitions and sufficient-penalty proofs are given in the Appendix.

A returned binary sample is interpreted as a path only when its selected
transitions form a single continuous source--target chain.  Branching,
cycles, disconnected selections, repeated nodes, and invalid termination
are rejected.  For each semantic condition, Dijkstra's algorithm establishes
the minimum path cost on the network or action multigraph.  A decoded path
is classified as optimal when it attains that cost, so equally costly paths
are accepted without requiring one canonical route.  Exhaustive enumeration
on tractable instances additionally checks feasible paths and QUBO energies.
The eligible-network and single-action models are compared for equivalence
of feasible paths, optimal costs, and representation size. A structural audit
checks the one-to-one transition correspondence and cost preservation in all
48 observations; full path enumeration is not required for this isomorphism.
Recorded ``exact'' recovery metrics mean recovery of a primary-cost optimum,
not use of an exact optimization algorithm.

\subsection{Outcomes and Inference}

The evaluation distinguishes finding a feasible attack path from recovering
one with minimum cost.  At the call level, success means that at least one
returned sample represents a valid or optimal path, respectively.  This
measures whether a solver execution produces a usable solution.  At the
sample level, recovery is the fraction of returned samples that are valid or
optimal, with repeated samples counted according to their observed
frequencies.  The two levels therefore describe complementary aspects of
performance: obtaining a solution from a call and the quality of its sample
population.

Invalid-path categories describe decoding failures, while variable and
coupling counts describe representation requirements. The archived analysis
also includes best-of-$R$ sample summaries, decoded-cost regret,
minimum-energy sample status, and runtime; these secondary outcomes are not
used to establish the paper's principal conclusions.

Statistical comparisons preserve the pairing between adjacent semantic
conditions on the same topology. Only pairs of completed calls enter these
contrasts; a skipped condition removes its corresponding pair from that
comparison. Observations are first averaged within each
master topology, which serves as the independent experimental unit.
Uncertainty is estimated using 95\% percentile intervals from 10,000 bootstrap
replicates that resample whole topologies. Exact paired sign tests assess the
direction of the contrasts, omitting zero differences. Benjamini--Hochberg
correction controls the false discovery rate separately for validity and
optimality outcomes. Each family contains both call- and sample-level
outcomes across four adjacent contrasts and six heuristic--formulation strata
(48 tests per family). The bootstrap intervals are not multiplicity-adjusted
and need not agree with decisions from the adjusted sign tests.

Archived exploratory regressions comprise a topology-clustered linear
probability model and a binomial-logit model with a Gaussian topology
intercept. Their specification replaces unavailable energy gaps by zero and
includes only a steepest-descent backend indicator, grouping simulated
annealing and tabu search in the reference category. These limitations prevent
interpreting the models as estimates of separate backend effects or a
recovery mechanism; they are not used as evidence for the conclusions here.
With only six independent topologies, paired effect sizes receive greater
emphasis than significance alone. A nonsignificant contrast is not interpreted
as evidence that the conditions are equivalent.

\subsection{Threshold Remediation}

Let $z_k\in\{0,1\}$ indicate whether vulnerability $k$ is patched; its
contribution to remediation cost is $p_kz_k$. A solution is
feasible when its budget is respected and the residual shortest chain for
every protected pair is absent or has cost at least the declared threshold.
Small scenarios are solved by exhaustive remediation-set enumeration and an
independent mixed-integer model that repeatedly checks residual shortest paths
until convergence.  QUBO solutions are evaluated against the same budget,
security thresholds, and minimum patch cost established by the exact
references. All eight scenarios share the same source topology.

\section{Results}
\label{sec:results}

\subsection{Representation Equivalence and Path Recovery}

Representing each eligible network transition by a single vulnerability-labeled
action preserves the underlying optimization problem.  Across all 48
topology--density--pair observations, the eligible-network and single-action
models have identical projected feasible paths, primary costs, sets of optimal
paths, and QUBO sizes.  Thus, identifying the vulnerability associated with a
transition does not itself increase the representation size or alter the
minimum path cost.

Representation equivalence does not require identical heuristic recovery
counts. A shared seed controls random-number generation, but variable
relabeling can change a heuristic's sampling trajectory.

The recovery results distinguish this change in representation from changes
in cost and action availability (Table~\ref{tab:stage-results}).  Eligibility
and vulnerability labeling changed recovery only slightly, whereas exploit-cost
scalarization raised all four pooled recovery measures.  Introducing multiple
exploit alternatives left valid-path recovery approximately unchanged but
reduced the fraction of calls returning an optimal path from 0.726 to 0.649
and the fraction of optimal samples from 0.571 to 0.458.
These pooled rates describe completed calls. The paired estimates below
instead average within-topology contrasts over matched completed conditions,
giving each master topology equal weight. In particular, the pooled increase
in call validity from 0.750 to 0.762 coexists with a slightly negative paired
contrast because the multiple-action hop condition excludes six calls.

\begin{table}[t]
\caption{Path-recovery rates across the five semantic conditions, pooled over
heuristics and QUBO formulations.}
\label{tab:stage-results}
\centering
\small
\begin{tabular}{lrrrrr}
\toprule
Condition & Calls & Any valid & Any exact & Read valid & Read exact \\
\midrule
Full network & 288 & 0.729 & 0.698 & 0.570 & 0.549 \\
Eligible network & 288 & 0.733 & 0.694 & 0.577 & 0.550 \\
Matched action & 288 & 0.726 & 0.684 & 0.575 & 0.547 \\
Exploit cost & 288 & 0.750 & 0.726 & 0.589 & 0.571 \\
Multiple action & 282 & 0.762 & 0.649 & 0.588 & 0.458 \\
\bottomrule
\end{tabular}
\par\smallskip
\begin{minipage}{0.95\textwidth}
\footnotesize\textit{Note:} ``Any valid'' and ``Any exact'' give the fractions
of sampled calls returning at least one valid or minimum-cost path,
respectively.  ``Read valid'' and ``Read exact'' give the corresponding
fractions of returned samples.  Repeated samples retain their observed
frequencies; samples and calls within a topology are not independent
experimental units. The experiment planned 1,440 calls, completed 1,434, and
returned 14,340 samples. Six multiple-action hop calls were excluded before
sampling under the resource policy specified in the Appendix.
\end{minipage}
\end{table}

Across the six backend--formulation strata, mean multiple-action minus
single-action exploit-cost differences were
$-0.003$ for call validity, $-0.090$ for call exactness, $-0.012$ for read
validity, and $-0.123$ for read exactness.  Every call-exact stratum was
negative (range $-0.167$ to $-0.021$), as was every read-exact stratum (range
$-0.396$ to $-0.040$).  Mean resource increments were 227 variables and
125,491 couplings.  No adjacent recovery contrast was statistically significant after
Benjamini--Hochberg correction within the corresponding outcome family.
Given the six independent topologies, these differences are interpreted as
descriptive estimates.
Table~\ref{tab:multiplicity-inference} reports the optimal-recovery contrasts
and their uncertainty. Some unadjusted bootstrap intervals exclude zero,
while no multiplicity-adjusted sign test rejects its null hypothesis; the
intervals and tests use different procedures.

\begin{table}[t]
\caption{Paired differences in optimal recovery: multiple actions minus the
single-action exploit-cost condition. Differences are in proportions;
intervals are unadjusted 95\% topology-bootstrap percentile intervals.
Adjusted $p$-values use the sign-test optimality family.}
\label{tab:multiplicity-inference}
\centering
\small
\begin{tabular}{lllrrr}
\toprule
Heuristic & Formulation & Outcome & Difference & 95\% interval & Adjusted $p$ \\
\midrule
Simulated annealing & Edge-flow & Call & $-0.063$ & $[-0.104,-0.021]$ & 0.857 \\
 & & Sample & $-0.396$ & $[-0.440,-0.340]$ & 0.375 \\
Steepest descent & Edge-flow & Call & $-0.146$ & $[-0.292,-0.042]$ & 0.600 \\
 & & Sample & $-0.071$ & $[-0.110,-0.033]$ & 0.375 \\
Tabu search & Edge-flow & Call & $-0.021$ & $[-0.063,0.000]$ & 1.000 \\
 & & Sample & $-0.092$ & $[-0.154,-0.027]$ & 1.000 \\
Simulated annealing & Hop & Call & $-0.167$ & $[-0.333,-0.042]$ & 0.857 \\
 & & Sample & $-0.044$ & $[-0.104,-0.004]$ & 1.000 \\
Steepest descent & Hop & Call & $-0.083$ & $[-0.125,-0.042]$ & 0.600 \\
 & & Sample & $-0.040$ & $[-0.063,-0.019]$ & 0.375 \\
Tabu search & Hop & Call & $-0.063$ & $[-0.146,0.000]$ & 1.000 \\
 & & Sample & $-0.094$ & $[-0.163,-0.033]$ & 0.500 \\
\bottomrule
\end{tabular}
\end{table}

\begin{figure}[t]
\centering
\includegraphics[width=\linewidth]{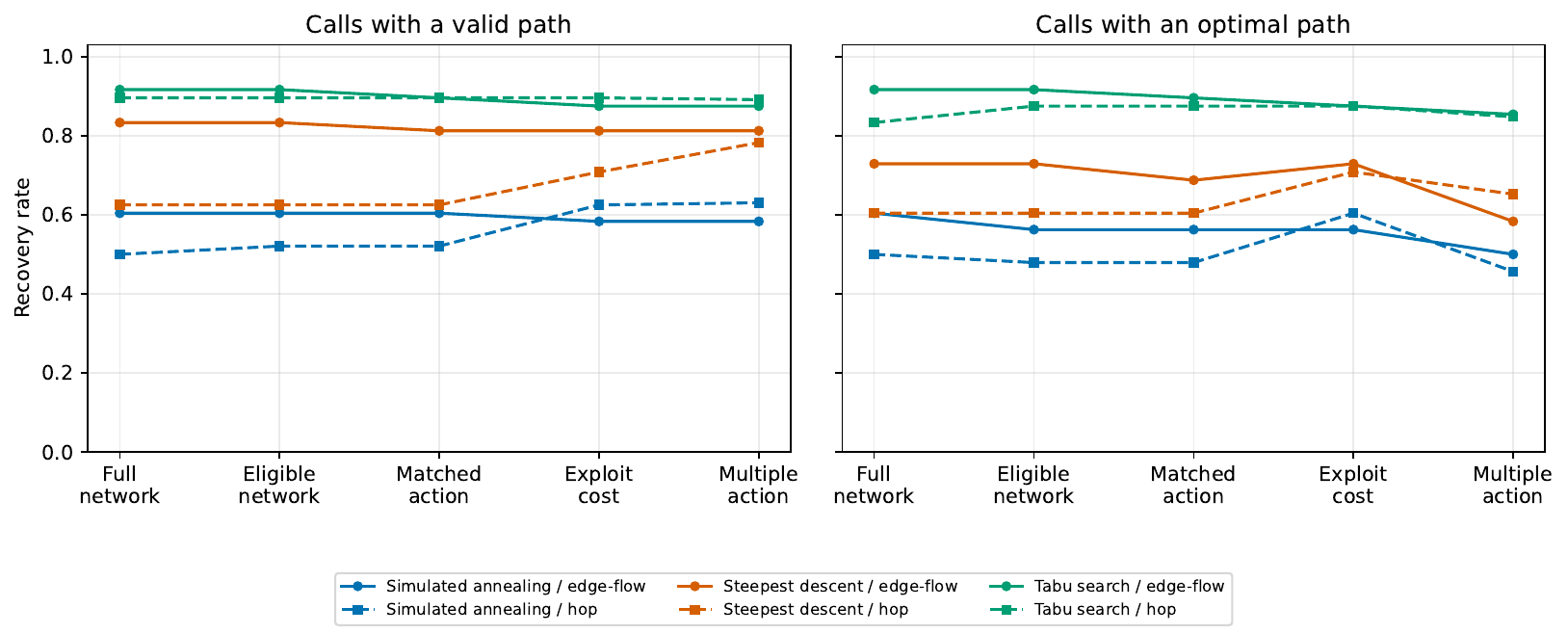}
\caption{Call-level path recovery across the five semantic conditions in the primary
experiment.  Each curve represents one heuristic and formulation.  The left
panel shows the fraction of completed calls returning at least one valid path;
the right panel shows the fraction returning at least one minimum-cost path.
Each curve uses 48 calls per condition, except multiple-action hop curves,
which use 46. Table~\ref{tab:stage-results} pools these denominators across
all six curves.}
\label{fig:call-recovery-size}
\end{figure}

\begin{figure}[t]
\centering
\includegraphics[width=\linewidth]{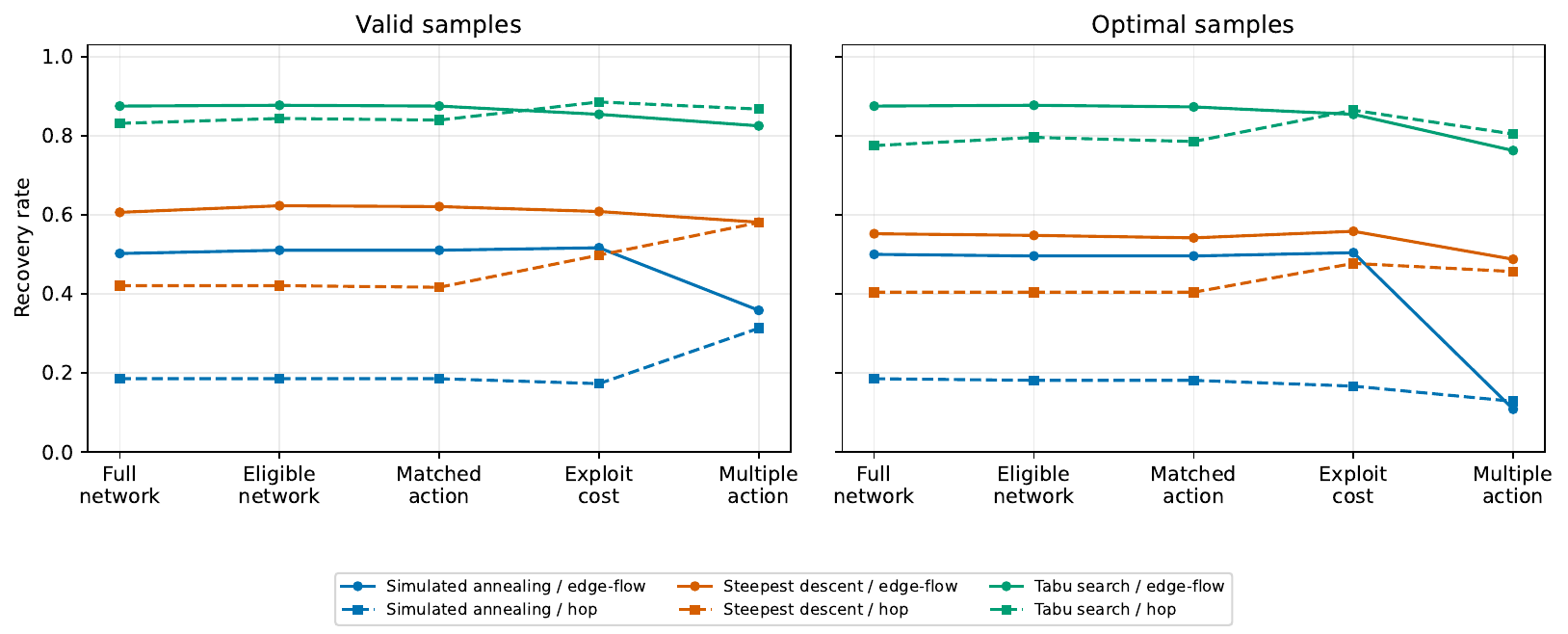}
\caption{Sample-level path recovery across the five semantic conditions.  The left
panel shows the fraction of returned samples that decode to valid paths;
the right panel shows the fraction that attain the minimum path cost.
Repeated samples contribute according to their observed frequencies.
Samples within a solver call are not independent experimental units.}
\label{fig:read-recovery-size}
\end{figure}

The remaining comparisons help distinguish changes in model size from
changes in recovery.  Adding vulnerability labels to eligible transitions
left variable and coupling counts unchanged, with mean recovery differences
ranging from $-0.010$ to $-0.002$.  Restricting transitions to exploit-eligible
edges reduced model size by an average of 12 variables and 4,188 couplings.
Incorporating exploit costs was instead associated with higher recovery:
the proportions of calls returning a valid or optimal path increased by
2.4 and 4.2 percentage points, respectively.  These descriptive differences
are calculated from paired observations within each topology.
Figures~\ref{fig:call-recovery-size} and \ref{fig:read-recovery-size} show how
recovery across the five conditions varies with the heuristic and formulation.

\subsection{Heuristics, Formulations, and Resources}

Across the five conditions, tabu search achieved the highest recovery rates
among the evaluated heuristics: pooled any-valid and
any-exact rates were 0.896 and 0.892 for edge-flow and 0.895 and 0.861 for hop.
The corresponding simulated-annealing rates were 0.596/0.558 and
0.559/0.504; steepest descent achieved 0.821/0.692 and 0.672/0.635.  These rates apply to the specified sampling settings.

\begin{figure}[t]
\centering
\includegraphics[width=\linewidth]{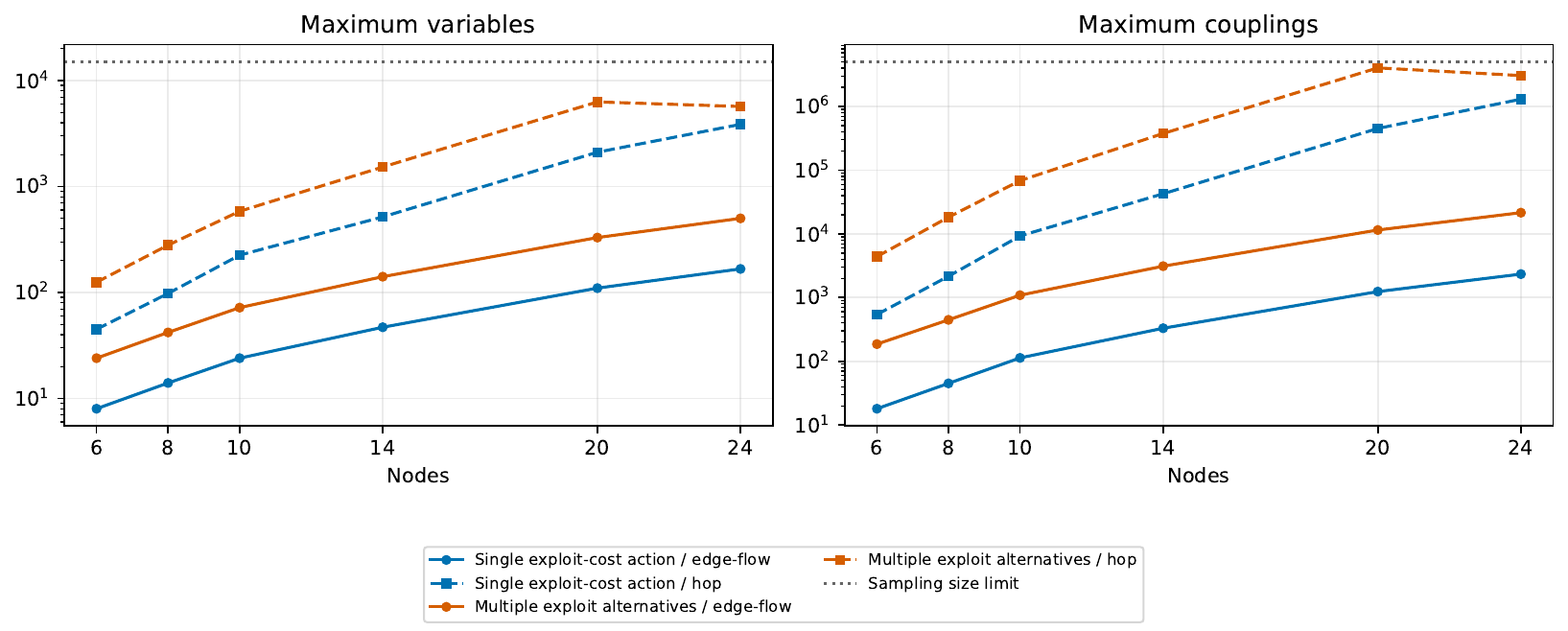}
\caption{Maximum QUBO sizes among sampled models at each node count for single
exploit-cost actions and multiple exploit alternatives.  Dashed curves denote
hop and solid curves denote edge-flow formulations.  Dotted horizontal lines
indicate the sampling resource limits.}
\label{fig:resource-growth}
\end{figure}

The largest sampled multiple-action edge-flow QUBO contained 501 variables and 21,516 couplings;
the largest completed multiple-action hop QUBO contained 6,289 variables and 4,033,146
couplings.  Invalid samples are included in the recovery-rate denominators.  The most
frequent causes of invalidity were unassigned hop positions (2,207 occurrences), empty
selection (1,412), cycles (1,160), and multiple actions for one compromise
(336).  Figure~\ref{fig:resource-growth} compares representation sizes across
graph sizes for the single-action exploit-cost and multiple-action conditions.

Lower optimal-path recovery under action multiplicity accompanied larger
representations and additional alternative paths. The archived regression
dataset contains 570 completed calls from the exploit-cost and multiple-action
conditions. Energy-gap and ground-state measurements are unavailable for 252
of these calls; the archived models replace the missing gaps with zero.
All 318 available ground-state degeneracy measurements equal one, so there is
no observed variation with which to evaluate a degeneracy--recovery
association. Together with the backend coding described in the methodology,
these limitations preclude using the regression outputs to establish a
recovery mechanism. The paired recovery contrasts and representation counts
provide the descriptive evidence reported here.

\subsection{Sensitivity, External Case, and Remediation}

The sensitivity experiment crosses sufficient and half-scale penalties with
geometric and automatically selected sampling schedules, using one read and
five sweeps per call. Each profile contains 20 calls, pooled over the five
conditions, two formulations, and two heuristics. Under either schedule
setting, the half-scale profile produced 18 invalid samples and no optimal
samples; the sufficient-penalty profile produced 14 invalid samples and four
optimal samples. Thus invalidity was 90\% versus 70\% in this diagnostic.
The profiles have identical aggregate counts across schedule settings, and
the schedule parameter applies only to simulated annealing. One topology
does not support general rankings of schedules or penalty choices.

The external 3CX evaluation comprises 180 calls and 3,600 samples.  At least one
valid chain appeared in 60.0\% of calls and an exact chain in 52.8\%; read-level
rates were 34.0\% valid and 28.4\% exact.  The eligible-network and single-action representations were equivalent in
all six comparisons, and 30 exact path references were computed.  These
results demonstrate application to a published topology under synthetic
costs; one source graph cannot establish performance across operational
networks.

Mixed-integer optimization and exhaustive enumeration agreed on the minimum
patch cost in all eight remediation scenarios.  Constraint generation
converged in two solves per scenario, and all QUBOs satisfied the
representation-size limits.  Of 16
fixed-effort calls (1,600 reads), 13 contained a valid remediation and 12
contained an exact one; 209 reads were valid and 155 exact.  These scenarios include vulnerabilities shared across actions and two
protected source--target pairs.  Patch selection therefore couples several
attack paths, requiring a joint optimization beyond a single shortest-path
calculation.

\section{Discussion}
\label{sec:discussion}

The results show that the effect of vulnerability information depends on how
it enters the path model.  Assigning one labeled action to each eligible
transition at the same cost preserves feasible paths, optimal costs, and QUBO
size.  Labels therefore make the vulnerability behind each compromise explicit
without changing the underlying optimization problem.  Restricting transitions
to exploit-eligible edges changes which routes are available, whereas exploit
costs change how those routes are evaluated.  Keeping these changes separate
allows the observed recovery differences to be interpreted more precisely than
a direct comparison between network-only and vulnerability-action models.

Introducing multiple exploit alternatives produces a different pattern:
valid-path recovery changes little, while optimal-path recovery is lower
under the tested sampling settings. This distinction matters because
a continuous attack chain need not be the least costly chain available to an
attacker.  The reduction in optimal recovery across all heuristic--formulation
combinations coincides with larger representations and greater competition
among alternative paths. These observations describe changes in representation
and recovery; they do not identify a causal search mechanism.
The recovery contrasts are not statistically significant
after correction for multiple comparisons, so their consistent direction
should be interpreted as a descriptive finding rather than a general effect.

The resource comparison helps explain why formulation choice matters even
when the path objective is unchanged.  The hop encoding makes order explicit
by assigning actions to path positions, increasing the number of variables
and interactions as the action catalog expands.  Edge-flow represents the
same simple-path problem more economically in these experiments.  It is
therefore the more compact option for the additive path objective studied
here, while hop may be useful when additional constraints depend on the
position or order of actions.  Solver choice also affects recovery: tabu
search performs best under the evaluated settings, although this ranking does
not establish superiority under different parameter choices or sampling
efforts.

The 3CX evaluation extends the analysis to a published attack graph and shows
that the formulations can recover valid and optimal chains beyond the
generated topologies.  Its contribution is an external application of the
model, rather than a calibration of operational risk. Entity and vulnerability
identities originate from the published dataset, while the conversion also
includes explicitly marked narrative-inferred transitions, synthetic costs,
and additional synthetic action alternatives. The recovered paths describe that graph
under the assigned costs, rather than predicting which attacks would occur
in practice.

Remediation gives the path model a further role in defensive decision making.
A patch can disable actions on several attack paths and for multiple protected
pairs, so patch selection requires coordinating decisions across the graph.
Shortest-path computation determines whether a proposed patch set achieves
the required protection thresholds, while the remediation optimization selects
a minimum-cost set within the budget.  Agreement between exhaustive and
mixed-integer references establishes the comparison target for the QUBO
solutions.  Recovery of optimal patch sets demonstrates feasibility for the
tested scenarios, while leaving the scalability of the coupled defender
problem open.

\subsection{Scope and Limitations}

The principal statistical limitation is the number of independent topologies.
The six synthetic master graphs support paired comparisons, but additional
source--target pairs, densities, solver calls, and samples remain observations within
those graphs.  They do not provide independent replications of the topology.
The external evaluation uses one published graph, and the penalty-sensitivity
experiment uses one topology; neither establishes performance across broader
network populations. Graph size and master-topology identity are confounded
because there is only one master topology per size. The eight remediation
scenarios derive from one six-node topology and therefore do not provide
eight independent network replications. Furthermore, the six pre-sampling
exclusions occur in the largest multiple-action hop models. Completed-call
recovery rates describe the sampled subset, not unconditional success over
all planned calls; the paired contrasts also omit unavailable pairs.

The security interpretation is also limited by the model assumptions.
Synthetic exploit and patch costs permit controlled comparisons but are not
empirical estimates of attacker effort, exploitation probability, or
remediation cost.  Applying the model to operational decisions would require
calibrating these quantities for the relevant assets and threat conditions.
The action-chain abstraction additionally omits AND/OR prerequisites and the
accumulation of privileges across compromises.  Concurrent actions and changes
in topology would require further extensions.

Finally, all sampled QUBO solutions were obtained using classical local
heuristics.  The study evaluates the correctness and resource requirements of
the encodings and their recovery behavior under those heuristics.  It does not
evaluate physical quantum hardware or the overheads of mapping these models
to a quantum processor.  Accordingly, the results provide a baseline for such
comparisons, without establishing a computational advantage over specialized
classical path algorithms or quantum advantage over classical optimization.

\section{Conclusion}
\label{sec:conclusion}

This study examined how vulnerability information affects QUBO models for
attack-path analysis and defensive remediation.  The matched comparisons
show that assigning one vulnerability-labeled action to each eligible
network transition at the same cost preserves feasible paths, optimal costs,
and representation size.  Introducing multiple exploit alternatives was
associated with lower optimal-path recovery, although valid-path recovery
remained approximately unchanged.  The accompanying increase in variables
and couplings was particularly pronounced for the hop formulation, making
edge-flow the more compact encoding for the additive path problem studied
here.  These findings distinguish the interpretability provided by
vulnerability labels from the computational demands introduced by additional
action choices.

The external 3CX evaluation demonstrates application to a published attack
graph under synthetic costs.  The remediation extension then uses attack-path
constraints to select patches within a budget, accounting for vulnerabilities
shared across actions and protected pairs.  Against exhaustive and
mixed-integer references, the tested QUBO heuristics recovered an optimal patch set in 12 of 16 calls
across eight scenarios.  This establishes feasibility at the tested scale
while leaving performance on larger defensive problems open.

The recovery differences remain descriptive: none of the adjacent
semantic-condition recovery contrasts was statistically significant after
correction for multiple comparisons, and the small number of independent
topologies limits generalization. Since the experiments use
classical heuristics, they also provide no evidence of quantum computational
advantage.  Further evaluation should use larger independent topology samples
and empirically calibrated costs.  Extending the model with observed
vulnerability and quantum-risk information will require separating attacker
effort from security consequences and remediation cost, so that the resulting
paths and patch priorities have a clear operational interpretation.

\section{Code and Data Availability}
\label{sec:reproducibility}

The project repository is
\url{https://github.com/Kasymkhankhubiyev/Qubo-Graph}.
Reproducibility materials include experimental configurations, topology and
action data, exact path and remediation references, QUBO summaries, sampled
binary assignments and their frequencies, solver outcomes, and analysis tables.
The validation routines reconstruct feasibility, optimality, sample counts,
and paired contrasts from the recorded results.  The external 3CX dataset is
identified by its source DOI in Section~\ref{sec:experimental-design}.

The following paths are relative to \nolinkurl{data/experiments/} in the
repository. The primary dataset is
\nolinkurl{matched_action_remediation_resource_pilot/20260804T155424.408093Z-matched-action-remediation-resource-pilot/}.
Its \nolinkurl{analysis/matched-action-remediation/} subdirectory contains the
paired contrasts and archived exploratory analyses. The supplementary datasets
are \nolinkurl{matched_action_remediation_sensitivity_smoke/},
\nolinkurl{external_3cx/20260804-completion-smoke/}, and
\nolinkurl{matched_action_remediation_factorial_smoke/} for sensitivity,
external evaluation, and remediation, respectively. Run and analysis manifests
record artifact checksums. The submission figure manifest also records the
primary call-file checksum and checksums of all three plotted PDFs.

The primary run manifest records commit
\nolinkurl{03582efe13f3fe76a7e10ce8ea0e634304c2e142} with a modified worktree
and source-tree checksum computed with the 256-bit Secure Hash Algorithm
(SHA-256):
\nolinkurl{13bd0d7382229f3dc152fc2711dfa929908e171dc6854535bfb8a28a1a712077}.
Thus the commit alone does not identify the executed source state. Stored
configuration, \nolinkurl{source-pyproject.toml}, and
\nolinkurl{source-uv.lock} snapshots document the run settings and dependency
versions; no separate immutable public release is identified here.

\appendix
\section{Formal Definitions and Proofs}
\label{app:formal}

\subsection{Path Objectives}

\begin{definition}[Network-path objective C1]
\label{def:c1}
Let $G=(V,E)$ be a directed graph with source $s$, target $t$, and positive
integer edge costs $w_{uv}$.  For a directed $s$--$t$ path $P$,
\begin{equation}
    C_1(P)=\sum_{(u,v)\in P}w_{uv}.
\end{equation}
C1 asks for an $s$--$t$ path minimizing $C_1(P)$.
\end{definition}

\begin{definition}[Vulnerability-action objective C2]
\label{def:c2}
For every eligible transition, let $a=(u,v,k)\in\Aset$ denote the action that
uses vulnerability $k$ to compromise $v$ from $u$.  Its positive integer cost
is
\begin{equation}
    c_a=c_a^{\mathrm{network}}+c_a^{\mathrm{effort}}
        +c_a^{\mathrm{time}}+c_a^{\mathrm{detection}}
        +c_a^{\mathrm{resource}}.
\end{equation}
C2 asks for a continuous ordered action chain from $s$ to $t$ minimizing
$\sum_a c_a$.  The source is controlled before the first action, and the last
real action compromises the target.
\end{definition}

Because all costs are positive and the model has no ordered logical
prerequisites, Dijkstra's algorithm on $G$ and on the action multigraph
supplies exact C1 and C2 reference costs.  An exact result need only attain the
minimum cost; it need not reproduce one canonical path.

\subsection{C1 QUBO Definitions}

The primary experiment uses a shared transition builder for all five
conditions. Network edges in the full- and eligible-network conditions are
treated as labeled transitions and receive the same presolve, count
augmentation, feasible-reference penalty, and hop-bound rules as actions.
The following C1 definitions describe these evaluated encodings. The primary
objective remains $C_1$; augmentation only resolves ties. The alternative
unaugmented edge-flow baseline with penalty $\sum w_{uv}+1$ is not the
encoding sampled in this matched experiment.

\begin{definition}[C1 edge-flow QUBO]
\label{def:c1-edge}
After the shared presolve in Definition~\ref{def:c2-presolve}, let $E'$ be
the retained non-self-loop edges and create $x_{uv}\in\{0,1\}$ for each.
Let $U$, $h_U$, and $\epsilon$ be the constructive-reference cost, length,
and count augmentation defined there, using edge weights as primary costs.
Write
\begin{align}
H_{\mathrm{cost}}^{(1)}(x)
    &=\sum_{(u,v)\in E'}(w_{uv}+\epsilon)x_{uv},\\
H_{\mathrm{flow}}^{(1)}(x)
    &=\sum_{v\in V}
      \left(
      \sum_{j:(v,j)\in E'}x_{vj}
      -\sum_{i:(i,v)\in E'}x_{iv}-b_v
      \right)^2,
\end{align}
where $b_s=1$, $b_t=-1$, and $b_v=0$ otherwise.  The QUBO energy is
\begin{equation}
\begin{aligned}
H_{\mathrm{edge}}^{(1)}(x)
  &=H_{\mathrm{cost}}^{(1)}(x)
   +\lambda_{\mathrm{flow}}H_{\mathrm{flow}}^{(1)}(x),\\
\lambda_{\mathrm{flow}}
  &=\frac{U+\epsilon h_U}{2}+0.5.
\end{aligned}
\label{eq:c1-edge}
\end{equation}
\end{definition}

Under exact minimization, positive costs remove cycles from a feasible binary
unit flow, leaving a minimum-cost $s$--$t$ path.  A heuristic sample is still
decoded strictly to reject branching, disconnected, cyclic,
boundary-violating, and ambiguous selections.

\begin{definition}[C1 ordered-hop QUBO]
\label{def:c1-hop}
Create transition-position variables $z_{a,p}$ for every canonical network
transition $a$ and position $p=0,\ldots,L-1$, together with one zero-cost
inactive terminal marker per position.  Nonnegative penalties enforce one label per
position, a first transition leaving $s$, endpoint continuity, arrival at
$t$, an inactive suffix, and no repeated nonterminal node.  Thus every simple
$s$--$t$ path of length at most the shared safe bound $L$ has a zero-penalty
representation and shorter paths do not require a target self-loop.
Use the explicit hop Hamiltonian in Definition~\ref{def:c2-hop}, replacing
actions by retained network transitions and $c_a$ by $w_{uv}$. Derive $U$,
$w_{\min}$, $L$, and $\epsilon$ from those transition costs. The sampling
penalty is $\lambda_{\mathrm{hop}}=U+\epsilon h_U+0.5$.
\end{definition}

\subsection{Shared Presolve and Primary-Cost Preservation}

\begin{definition}[Shared presolve and augmented cost]
\label{def:c2-presolve}
Remove actions entering $s$ or leaving $t$, and retain only actions whose
endpoints are both forward-reachable from $s$ and reverse-reachable to $t$.
Let a deterministic breadth-first search provide a feasible action chain with
primary cost $U$ and length $h_U$.  For $n>2$ and $0\leq\alpha<1$, define
\begin{equation}
    \widetilde c_a=c_a+\epsilon,\qquad
    \epsilon=\frac{\alpha}{n-2},
\end{equation}
with $\alpha=0.5$ in the experiment.
The same construction applies to network transitions in C1 by substituting
their primary edge weights for action costs. The shared experiment has
$n\geq6$ and distinct source and target nodes.
\end{definition}

Two simple paths differ in action count by at most $n-2$, so the augmentation
changes their ordering by less than one integer primary-cost unit.  It may
break equal-cost ties toward fewer actions but cannot prefer a
higher-primary-cost path.

\subsection{C2 Action Edge-Flow QUBO}

\begin{definition}[C2 action edge-flow QUBO]
\label{def:c2-edge}
Create $q_a\in\{0,1\}$ for each retained action and define
\begin{equation}
    r_v(q)=
    \sum_{a:\operatorname{tail}(a)=v}q_a
    -\sum_{a:\operatorname{head}(a)=v}q_a-b_v.
\end{equation}
The action edge-flow energy is
\begin{equation}
H_{\mathrm{edge}}^{(2)}(q)
 =\sum_{a\in\Aset}(c_a+\epsilon)q_a
  +\lambda_{\mathrm{flow}}\sum_{v\in V}r_v(q)^2.
\label{eq:c2-edge}
\end{equation}
\end{definition}

\begin{proposition}[Sufficient action-flow penalty]
\label{prop:action-flow}
If $2\lambda_{\mathrm{flow}}>U+\epsilon h_U$, every ground state of
\eqref{eq:c2-edge} decodes to a simple source-to-target action chain with
minimum primary cost; the action-count augmentation affects ties only.
\end{proposition}

\begin{proof}
Every residual is integral and $\sum_v r_v=0$.  A nonzero residual vector
therefore has squared norm at least two, so every infeasible state costs more
than the known feasible chain.  A feasible binary unit flow decomposes into an
$s$--$t$ path and zero or more cycles.  Removing any cycle preserves flow and
strictly reduces the positive augmented cost.  A ground state is consequently
a simple path.  The bound on $\epsilon$ preserves primary integer-cost
ordering.
\end{proof}

The experiment uses
\begin{equation}
    \lambda_{\mathrm{flow}}
      =\frac{U+\epsilon h_U}{2}+\delta,\qquad \delta=0.5,
\end{equation}
before positive max-absolute-bias normalization.

\subsection{C2 Ordered Action-Hop QUBO}

\begin{definition}[C2 ordered action-hop QUBO]
\label{def:c2-hop}
Let $w_{\min}$ be the smallest retained action cost and set
\begin{equation}
    L=\min\{n-1,\lfloor U/w_{\min}\rfloor\}.
\end{equation}
For every position $p=0,\ldots,L-1$, create a variable $z_{a,p}$ for each
retained real action and one inactive terminal marker.  Nonnegative penalties enforce
one label per position, a first action leaving $s$, endpoint continuity,
termination at $t$, an inactive suffix, and no repeated nonterminal
compromise target.  Only real actions carry $\widetilde c_a$.
For an explicit Hamiltonian, let $\mathcal B=\Aset\cup\{\bot\}$, where
$\bot$ is the inactive marker, and let $\mathcal T$ contain the allowed
consecutive-label pairs: $(a,b)$ for real actions with
$\operatorname{head}(a)=\operatorname{tail}(b)$ and
$\operatorname{head}(a)\ne t$; $(a,\bot)$ when
$\operatorname{head}(a)=t$; and $(\bot,\bot)$. Define
\begin{align}
P_{\mathrm{one}}&=\sum_{p=0}^{L-1}
  \left(1-\sum_{a\in\mathcal B}z_{a,p}\right)^2,\\
P_{\mathrm{boundary}}&=z_{\bot,0}
  +\sum_{a:\operatorname{tail}(a)\ne s}z_{a,0}
  +\sum_{a:\operatorname{head}(a)\ne t}z_{a,L-1},\\
P_{\mathrm{continuity}}&=\sum_{p=0}^{L-2}
  \sum_{(a,b)\notin\mathcal T}z_{a,p}z_{b,p+1},\\
P_{\mathrm{repeat}}&=\sum_{v\ne t}\sum_{0\leq p<q<L}
  \sum_{\substack{a,b\in\Aset:\\
    \operatorname{head}(a)=\operatorname{head}(b)=v}}
    z_{a,p}z_{b,q}.
\end{align}
The boundary sums range over real actions and the continuity sum over
$\mathcal B^2\setminus\mathcal T$. Actions entering $s$ or leaving $t$
have already been removed by presolve. The energy is
\begin{equation}
\begin{split}
H_{\mathrm{hop}}^{(2)}={}&
 \sum_{p=0}^{L-1}\sum_{a\in\Aset}(c_a+\epsilon)z_{a,p}\\
 &+\lambda_{\mathrm{hop}}(P_{\mathrm{one}}+P_{\mathrm{boundary}}\\
 &\hspace{4.2em}+P_{\mathrm{continuity}}+P_{\mathrm{repeat}}).
\end{split}
\label{eq:c2-hop}
\end{equation}
The experiment uses $\lambda_{\mathrm{hop}}=U+\epsilon h_U+0.5$ before
normalization. The bound $L$ retains every primary-cost optimum because an
optimal path costs at most $U$ and each action costs at least $w_{\min}$;
it need not retain higher-cost feasible paths of greater length.
\end{definition}

\begin{proposition}[Sufficient action-hop penalty]
\label{prop:action-hop}
If $\lambda_{\mathrm{hop}}>U+\epsilon h_U$, every ground state of the ordered
action-hop model decodes to a simple source-to-target action chain with minimum
primary cost.
\end{proposition}

\begin{proof}
Every infeasible assignment activates at least one nonnegative unit penalty
and therefore costs more than the known feasible chain.  A zero-penalty
assignment begins at $s$, follows continuous applicable actions, reaches $t$,
and then uses only inactive markers; the no-repeat condition makes its real prefix
simple.  Minimizing real-action cost over these assignments and applying the
primary-cost-preservation argument yields a minimum-primary-cost chain.
\end{proof}

Before solver reductions, the action-hop formulation uses exactly $L(|\Aset|+1)$ variables.

\begin{definition}[QUBO energy convention]
\label{def:qubo-output}
Both C2 formulations use the upper-triangular convention
\begin{equation}
E(x)=\mathrm{offset}+\sum_{i\leq j}Q_{ij}x_ix_j.
\end{equation}
Constants remain in the offset.  Positive normalization scales coefficients,
offset, and penalties together and therefore preserves minimizers.
\end{definition}

Six multiple-action hop calls were excluded before sampling by the resource
policy. They correspond to two 24-node models, each scheduled for three
heuristics, with 10,373 variables and 9,898,636 couplings, and 11,044 variables
and 11,451,534 couplings, respectively. Both exceed the 5,000,000-coupling
limit; the other limits are 15,000 variables and 5,015,000 total terms.
Consequently, 1,434 of 1,440 planned calls completed and returned 14,340
samples. Recovery rates use completed calls, and paired contrasts use only
available completed pairs.

\subsection{Threshold-Remediation QUBO}

Let $z_k\in\{0,1\}$ denote remediation of vulnerability $k$ at positive
integer cost $p_k$, and let the budget $B$ be a nonnegative integer. Let
$\mathcal P$ contain every simple protected-pair path below its required
threshold, and let $K(P)$ contain the distinct remediable vulnerability
identifiers on path $P$. Binary witness $y_{Pk}$ selects one patched
vulnerability in $K(P)$. Binary slack $s_j$ represents unused budget.
Starting with represented range $R=0$ and index $j=0$, choose
$w_j=\min\{2^j,B-R\}$, update $R\leftarrow R+w_j$, and continue until
$R=B$; for $B=0$ no slack variables are needed. These weights represent
every integer slack from zero to $B$. A feasible patch set is required before
the QUBO is built, and every path in $\mathcal P$ must have a nonempty
$K(P)$. The remediation Hamiltonian is
\begin{align}
\begin{split}
H={}&\sum_k p_kz_k\\
&+A\sum_{P\in\mathcal P}\Biggl[
  \left(1-\sum_{k\in K(P)}y_{Pk}\right)^2\\
  &\qquad\qquad+\sum_{k\in K(P)}y_{Pk}(1-z_k)
\Biggr]\\
&+A\left(\sum_kp_kz_k+\sum_jw_js_j-B\right)^2.
\end{split}
\label{eq:remediation}
\end{align}
With $A=\sum_kp_k+1$, any unit constraint violation costs more than the
largest possible patch-objective improvement.  Positive action costs make
hitting every path in $\mathcal P$ equivalent to raising each residual
shortest path to its threshold or disconnecting the pair.  Constants remain
in the offset, and strict decoding independently checks budget, witnesses,
residual shortest paths, and objective decomposition.

\section*{Funding}
The research and article processing charge were funded by the
Russian Science Foundation (project No.~25-11-20037,
\url{https://rscf.ru/project/25-11-20037/}).

\end{document}